\documentclass[11pt,epsfig]{article}
\def \la{\mathrel{\mathchoice   {\vcenter{\offinterlineskip\halign{\hfil
$\displaystyle##$\hfil\cr<\cr\sim\cr}}}
{\vcenter{\offinterlineskip\halign{\hfil$\textstyle##$\hfil\cr
<\cr\sim\cr}}}
{\vcenter{\offinterlineskip\halign{\hfil$\scriptstyle##$\hfil\cr
<\cr\sim\cr}}}
{\vcenter{\offinterlineskip\halign{\hfil$\scriptscriptstyle##$\hfil\cr
<\cr\sim\cr}}}}}
\def \ga{\mathrel{\mathchoice   {\vcenter{\offinterlineskip\halign{\hfil
$\displaystyle##$\hfil\cr>\cr\sim\cr}}}
{\vcenter{\offinterlineskip\halign{\hfil$\textstyle##$\hfil\cr
>\cr\sim\cr}}}
{\vcenter{\offinterlineskip\halign{\hfil$\scriptstyle##$\hfil\cr
>\cr\sim\cr}}}
{\vcenter{\offinterlineskip\halign{\hfil$\scriptscriptstyle##$\hfil\cr
>\cr\sim\cr}}}}}

\usepackage{epsfig}
\title{\bf Dwarf Galaxies in the Local Group\\ \medskip
and in the Local Volume}

\author{Eva~K.~Grebel\\
\vspace{0.1cm}\\
\normalsize  Max-Planck-Institut f\"ur Astronomie, K\"onigstuhl 17, D-69117
                 Heidelberg, Germany\\
}
\date{}
\begin{document}
\maketitle
\def\bull{\vrule height .9ex width .8ex depth -.1ex}
\makeatletter
\def\ps@plain{\let\@mkboth\gobbletwo
\def\@oddhead{}\def\@oddfoot{\hfil\tiny
``Dwarf Galaxies and their Environment'';
Bad Honnef, Germany, 23-27 January 2001; Eds.{} K.S. de Boer, R.-J.Dettmar, U. Klein; Shaker Verlag}%
\def\@evenhead{}\let\@evenfoot\@oddfoot}
\makeatother

\begin{abstract}\noindent
After summarizing the characteristics of different types of dwarf
galaxies I briefly review our current state of knowledge of dwarf
galaxy evolution in the Local Group, for which we now have a fairly
detailed although by no means comprehensive picture.  All Local Group
dwarfs studied to date contain an old population, though its fraction varies
considerably. The majority
of the dwarf companions of the Milky Way shows evidence for a common
epoch of ancient star formation.  Spatial variations in
star formation are frequently observed in many dwarf galaxies
in the Local Group and beyond.  These spatial variations range from a
seemingly stochastic distribution of star-forming regions in gas-rich,
high-mass dwarfs to radial gradients in low-mass dwarfs.  The global
mode of star formation may be either continuous with amplitude 
variations or episodic.  High-mass dwarf galaxies tend to form stars
over a Hubble time, whereas low-mass dwarfs eventually cease to form
stars, possibly aided by environmental effects.  Much less is
known about the content and properties of dwarf galaxies in the Local
Volume, which we are trying to remedy through a large observational
effort.  Dwarf galaxies in the Local Volume follow a similar trend
with absolute magnitude, mean metallicity, and central surface
brightness as the Local Group dwarfs do, and appear to be subject to
morphological segregation.  
\end{abstract}

\section{Introduction}

Galaxies cover a vast range of types and properties.  The by far most
frequent type of galaxy are dwarf galaxies, i.e., galaxies with low mass, 
low luminosity, low metallicity, and small in size.  The terminological
distinction between dwarf galaxies and ``normal'' or giant galaxies is
poorly defined and is used in different ways by different authors.  For
the purpose of this paper we shall use the absolute $V$-band magnitude
as a simple distinguishing criterion and consider all galaxies fainter 
than $M_V = -18$ mag as dwarf galaxies.  A potentially more meaningful
criterion would be mass rather than luminosity since dwarf galaxies 
have a range of different mass-to-light ratios.  Such a definition would,
for instance, consider all galaxies with masses of 1/10 or 1/100 of 
$M^*$ as dwarf galaxies.  However, since {\em masses} are unknown for
most dwarf galaxies while luminosities are a directly measurable 
observable, we will use the luminosity criterion here.

In hierarchical structure formation scenarios dwarf galaxies (or more
generally: small dark matter halos) are important building blocks of 
more massive galaxies and should have been even more numerous in the
early Universe.  Accordingly, the dwarfs we observe today are considered
survivors
of an initially much richer population.  Interestingly, at the low-mass end
we observe fewer dwarf galaxies today than the number of predicted  
surviving dark matter halos (e.g., Klypin et al.\ 1999).  This may imply 
that we are either missing
a large fraction of low-mass galaxies, that these halos do not have 
optically detectable counterparts, or that the scenarios are wrong.

Today's dwarf galaxies are the closest currently known counterparts of
the early galaxy-forming fragments.  
Their star formation histories may hold the clue to understanding the
conditions and evolution of the early Universe.  Especially low-mass 
dwarf galaxies tend to have experienced little enrichment and few star
formation episodes, i.e., they are ``simple systems'' as compared to 
more massive galaxies.  These properties make the study of their evolutionary
histories particularly interesting.  They are also interesting objects
from the point of view of stellar evolution since here we can study star
formation histories in low-metallicity environments, investigate mass
functions in low-mass, low-density environments, study the impact of 
neighboring massive galaxies on star formation, find out about preconditions
for star cluster formation, etc. 

These studies give the most accurate and reliable results where stellar
populations can actually be resolved into individual stars.  This makes
nearby dwarf galaxies in the Local Group and Local Volume excellent targets.

The Local Group is the small group of galaxies around the Milky Way and 
M31.  Within a zero-velocity surface of 1.2 Mpc (Courteau \& van den Bergh
1999), the Local Group contains
36 known or suspected member galaxies.  Since the orbits of these
galaxies are still unknown,
it is uncertain whether all of them are members, although one usually 
assumes that dwarf galaxies within $\sim 200$ kpc of a more massive galaxy
are bound satellites of that galaxy.  On the other
hand, we may be missing quite a few low-luminosity members that have yet
to be discovered.  For example, recent searches three to four years ago
uncovered four new, faint Local Group dwarfs (Armandroff, Davies, \& Jacoby 
1998; Karachentsev \& Karachentseva 1999; Armandroff, Jacoby, \& Davies
1999; Whiting, Hau, \& Irwin 1999).
 
The Local Volume (and Local Void) comprises the volume with a radius 
of about 5 Mpc.  It includes galaxies with velocities $\le 500$ km s$^{-1}$
that are either part of small galaxy groups or field galaxies.  The exact
number of galaxies and their faint-end census in the Local Volume
remain unknown.  A recently completed all-sky search led to the detection
of $\sim 600$ dwarf and low-surface-brightness galaxies, whose angular
extent indicates that they may belong to the Local Volume (Karachentsev
et al.\ 2000a).  More than half
of these objects were new discoveries.  

\section{Morphological types of dwarf galaxies}

{\bf Dwarf spiral galaxies} are found among S0, Sa, Sb, Sc, and Sd galaxies.
They have central surface brightnesses of $\mu_V \ga 23$ mag
arcsec$^{-2}$, H\,{\sc i} masses of $M_{\rm HI} \la 10^9$ M$_{\odot}$,
and usually large mass-to-light rations inferred from rotation curves
(e.g., Schombert et al.\ 1995; Matthews \& Gallagher 1997).
Dwarf spirals are at the high-mass end of the dwarf galaxy distribution.
Early-type dwarf spirals show rotation curves typical for
rotationally supported exponential disks, while late-type dwarf spirals
are slow rotators or exhibit solid-body rotation. 
Dwarf spirals show slow continuous star formation.  Later types tend
to be more metal-poor and gas-poor than earlier types (McGaugh 1994).  
Very late-type spirals may be in transition to irregulars.  Dwarf spirals 
are found in clusters and in the field. 

{\bf Blue compact dwarf galaxies} (BCDs) are actively star-forming 
objects with centrally concentrated starbursts.  Both gas and stars
show pronounced central concentration, contributing to the compact
appearance, fueling the starburst, and leading to high central
surface brightnesses (typically $\mu_V \la 19$ mag arcsec$^{-2}$).  
Various subtypes of dwarf galaxies exhibit BCD characteristics
including  H\,{\sc ii} galaxies, blue amorphous galaxies, and certain 
types of low-mass Wolf-Rayet galaxies.  
The H\,{\sc i} masses of BCDs are $\le 10^9$ M$_{\odot}$ and may 
exceed the stellar mass.  Part of the very extended gas may be 
kinematically decoupled from the galaxies (van Zee, Skillman, \& 
Salzer 1998).  
In their inner parts BCDs show solid-body rotation.  BCDs are often
observed in relative isolation away from the dense centers of 
clusters or groups. 

{\bf Dwarf irregular galaxies} (dIrrs) are irregular in appearance
and appear dominated by scattered bright H\,{\sc ii} regions in the 
optical.  In H\,{\sc i} these gas-rich galaxies show a complicated
fractal pattern full of shells and clumps.  In the more massive dIrrs
the  H\,{\sc i} distribution
is usually much more extended than even the oldest stellar populations.
In some cases evidence for gas accretion has been observed (e.g., 
Wilcots \& Miller 1998).
In less massive dIrrs the H\,{\sc i} may be displaced with respect to
the stellar light distribution, either off-centered or showing a ring-like
structure with a central depression (e.g., Young \& Lo 1997a). 
DIrrs are characterized by $\mu_V \la 23$ mag
arcsec$^{-2}$, $M_{\rm HI} \la 10^9 M_{\odot}$,
and $M_{\rm tot} \la 10^{10} M_{\odot}$.  The more massive dIrrs 
show increasing chemical enrichment with decreasing age. 
In low-mass dIrrs gas and stars may show distinct spatial distributions 
and kinematics.  Solid-body rotation is common among the more massive
dIrrs, while low-mass dIrrs occasionally do not show measurable rotation.
DIrrs are found in clusters, groups, and in the field.  They exhibit 
little concentration toward massive galaxies.  Massive dIrrs may 
continue to form stars over a Hubble time (Hunter 1997).
Very low-mass, gas-poor dIrrs, on the other hand, may be evolving 
towards dwarf spheroidals.

{\bf Dwarf elliptical galaxies} (dEs) are spherical or elliptical in
appearance, compact, have high central stellar densities 
and are typically fainter than $M_V = -17$ mag.  Other characteristics
include $\mu_V \la 21$ mag arcsec$^{-2}$, $M_{\rm HI}
\la 10^8 M_{\odot}$, and $M_{tot} \la 10^9 M_{\odot}$.
They tend to be found close to massive galaxies,
usually have little or no detectable gas, and are often not rotationally
supported.  If gas is detected it may show an asymmetric distribution,
be less extended than the underlying stellar light, and may be kinematically
distinct (Young \& Lo 1997b; Sage, Welch, \& Mitchell 1998; Welch, Sage, \&
Mitchell 1998).  DEs may show pronounced nuclei that can contribute up to 20\% 
of the galaxy's light.  The fraction of these nucleated dEs (dE(N))
increases among more luminous dEs.
The surface density profiles of dEs and dE(N)s are best described by
S\'ersic's (1968) generalization of a de Vaucouleurs $r^{1/4}$ law
and exponential profiles (Jerjen et al.\  2000).

{\bf Dwarf spheroidal galaxies} (dSphs) are diffuse, gas-deficient,
low-surface-brightness dwarfs with very little central concentration.  
They are characterized by $M_V \ga -14$ mag, $\mu_V \ga 22$ mag arcsec$^{-2}$,
$M_{\rm HI} \la 10^5 M_{\odot}$, and $M_{tot} \sim 10^7 M_{\odot}$.
DSphs are the faintest, least massive galaxies known.  The upper limits
for their H\,{\sc i} are usually even below the amounts expected from
normal stellar mass loss.   DSphs are usually found
in close proximity of massive galaxies
and do not appear to be supported by rotation.  
Their velocity dispersions indicate the presence of a significant dark
component when virial equilibrium is assumed. 

The difference in the 
spatial distribution of dEs and dSphs on the one hand and dIrrs on the 
other hand is also known as {\em morphological segregation} or as the
morphology--density relation for dwarf galaxies and may be
a signature of environmental effects on galaxy evolution.  

{\bf Tidal dwarf galaxies} form from the debris
torn out of more massive galaxies during interactions and mergers.  In
contrast to other dwarf galaxies they do not contain dark matter
and may have high metallicities for their luminosity depending on the
evolutionary stage of the parent galaxy (Duc \& Mirabel 1998).  Masses,
sizes, gas content, etc. of tidal dwarfs depend critically on the 
conditions during the interaction and on the parent galaxy.  It is 
possible that certain dIrrs and dSphs are tidal dwarfs that formed
relatively early.  For instance, various authors suggested that the 
Milky Way dwarf companions may be part of two or more major debris streams
in polar orbits (e.g., Majewski 1994; Lynden-Bell \& Lynden-Bell 1995).  
Potential candidates for nearby tidal dwarf galaxies based on their 
intrinsic properties are discussed by Hunter et al.\ (2000).  

\section{Dwarf Galaxies in the Local Group}

The dwarf galaxies studied most extensively to date are dwarf galaxies 
within the Local Group.  Of the above mentioned types, the Local Group
contains dIrrs, dEs, and dSphs as well as galaxies that may be in 
transition from dIrrs to dSphs.  

The results from the detailed studies of Local Group dwarfs can be 
(incompletely) summarized as follows.  

Local Group dwarfs were found to vary widely in their
star formation histories, mean metallicity and enrichment histories,
times of their major star formation episodes, fractional distribution
of ages and subpopulations.  Indeed no two dwarf galaxies are alike,
not even if they are of the same morphological type or have similar
luminosities (Grebel 1997).  In part their properties appear to be 
correlated with
galaxy mass and with environment (proximity to massive galaxies).

In many Local Group dwarfs spatial variations in star formation 
history were detected.  In general, there is a tendency for the younger
populations to be more centrally concentrated (and possibly more enriched),
whereas older populations are more extended.  This may in part be an 
effect of dynamical evolution, but may also indicate that star formation
was able to continue over an extended period of time in the central regions
since these may have been able to retain gas longer (or may have been
rejuvenated by gas inflow in accretion events).  For a summary of 
population gradients in a number of Local Group dSphs, see Harbeck et
al.\ (these proceedings).  Harbeck et al.\ conclude that presence or
absence of these gradients does not show a clear correlation with
environment. 

A common property to all dwarf galaxies studied in sufficient detail so
far is the presence of an old population.  All claims of discoveries of
purely young galaxies have subsequently been refuted once deeper images or
data with better coverage of the outskirts of these galaxies became 
available.  Deep ground-based imaging of the ``halos'' of dIrrs led to
the detection of old red giant branches (e.g., Minniti \& Zijlstra 1996)
All dSphs and dEs for which  deep photometry was obtained revealed 
horizontal branches, a clear indicator of old populations.
The nearest candidate for a dwarf spiral, NGC\,3109, shows a presumably
old red giant branch in its outer regions (Minniti, Zijlstra, \& Alonso 1999).
Deep imaging of BCDs has revealed the 
presence of intermediate-age and possibly old populations (e.g., Lynds 
et al.\ 1998), as already inferred from integrated colors.  

The oldest populations in the majority of nearby dIrrs and dSphs are 
indistinguishable in age from the oldest globular clusters in the Galactic
halo and bulge (for a full list and references, see Grebel 2000).  This
indicates a common epoch of the oldest measurable star formation episode.
Relative ages were derived from deep photometry extending below the 
main-sequence turn-off.  In at least one dIrr (the Small Magellanic Cloud), 
however, significant star
formation appears to have started with a delay of 2--3 Gyr.
Data of sufficient depth and quality to carry out differential age-dating
of the oldest populations are so far only available for Milky
Way companions, i.e., dwarf galaxies within $\sim 300$ kpc.

Nearby dwarf galaxies typically show one of two modes of star formation: 
Continuous or episodic.  High-mass dwarf galaxies tend to form stars 
fairly continuously with amplitude variations of up to a factor of three.
They can continue to form stars over another Hubble time without exhausting
their gas supply.  They exhibit gradual enrichment, since many of the metals
produced in star formation episodes are recycled later on.  This mode of
star formation is observed in high-mass dIrrs and dwarf spirals.
Lower-mass dIrrs, dEs, and dSphs show continuous star formation with 
decreasing star formation rates.  In dEs and dSphs, only little 
gas or no gas at all is observed, and star formation has ceased at the
present time.  One dSph in the Local Group, Carina, shows episodic or
repetitive star formation:  This galaxy underwent star formation episodes
separated by 2 Gyr long episodes of quiescence.  Yet no significant
enrichment appears to have occurred if one interprets the narrowness of
the red giant branch as an indication of a narrow metallicity range. 
This may suggest that the later episodes took place with
accreted gas.  Alternatively, age and metallicity may conspire to form
one narrow red giant branch.  The resolution of this question requires
spectroscopy.
It is not yet understood what caused the periods of quiescence
and the renewed onset of star formation lateron.

DSphs around the Milky Way tend to have increasing fractions
of intermediate-age populations with increasing distance from
the Milky Way, suggestive of the impact of environmental effects:
Tidal and ram pressure stripping may have removed most of the gas from
nearby dSphs very early on, whereas more distant dSphs were able to retain
their star-forming material for longer periods of time (e.g., van den Bergh
1994).  
The two dSph galaxies with the shortest Galactocentric distances (Ursa Minor
and Draco) are essentially entirely old.  These two are also the two least 
massive dwarfs.  The second most massive dSph in the Local Group, Fornax,  
ceased to form stars only about 200 Myr ago (Grebel \& Stetson 1999).  
(Here the term ``mass'' assumes
that luminosity traces mass.)  Hence the star formation history depends on
parent galaxy mass.  There are a number of uncertainties
with this simplified picture though.  Firstly, we do not know the orbits
of the Milky Way companions and cannot tell whether their present-day
positions are typical for their past distances from the Milky Way.  We 
also do not know how close they came to the Milky Way in the past, and what
role interactions played.  Secondly, no fully satisfactory mechanism for
the gas loss has been found so far.  Simulations indicate that internal
effects such as supernovae winds are insufficient to rid a dwarf galaxy of
its gas (Mac Low \& Ferrara 1999).  Mayer et al.\ (2001) suggest that tidal
shocks during perigalactic passages close to a massive galaxy may convert
dIrrs into dSphs.  They show that their simulations can reproduce the observed
global properties of the Milky Way dSph companions as well as the 
morphology-density relation.  Taking this model one step further, we would
not expect to find isolated dSphs.  Thus thirdly, if environmental effects 
determine the evolution of low-mass dwarfs, the existence of the dSph
Tucana in the distant outskirts of the Local Group is hard to explain.  This
isolated dSph does not show evidence for gas or young or intermediate-age 
populations.

The M31 dSph companions, which span a similar
range of distances around M31 as the Milky Way dSphs around the Galaxy 
provide a valuable source for testing the ram pressure/tidal stripping
scenario.  Interestingly, these galaxies do not show evidence for the 
presence of young or intermediate-age populations as traced by extended
blue main sequences or pronounced concentrations of red clump stars.  
Our preliminary analysis shows little indication for evolutionary trends
and present-day distance to M31.  

We may be observing ongoing evolution from low-mass dIrrs to dSphs.  
Low-mass dIrrs show little or no rotation and have low current star
formation rates.  Their H\,{\sc i} is not necessarily concentrated on
the optical center of the dwarf.  The so-called
transition-type galaxies may be in an advanced phase of evolution from
dIrrs to dSphs.  The measured stellar velocity dispersion of the dIrr/dSph
galaxy LGS\,3 is comparable to that of dSphs (Cook et al.\ 1998).
Mateo (1998) argues that the three transition-type galaxies LGS\,3, Phoenix,
and GR\,8  lie on
the same branch as dSphs when plotting [O/H] or [Fe/H] versus
absolute magnitude.  The recent stellar radial velocity determination 
for Phoenix by Gallart et al.\ (2001) argues in favor of the H\,{\sc i}
detected in the vicinity of Phoenix being associated with this dIrr/dSph
galaxy, which experienced star formation as recently as 100 Myr ago
(Holtzman et al.\ 2000).
The dSph Fornax, which still formed stars
some 200 Myr ago (Grebel \& Stetson 1999)
while being devoid of gas at present (Young 1999), may be a former dIrr that
only recently finished its transition to a dSph.  

\begin{figure}[ht]
\epsfig{file=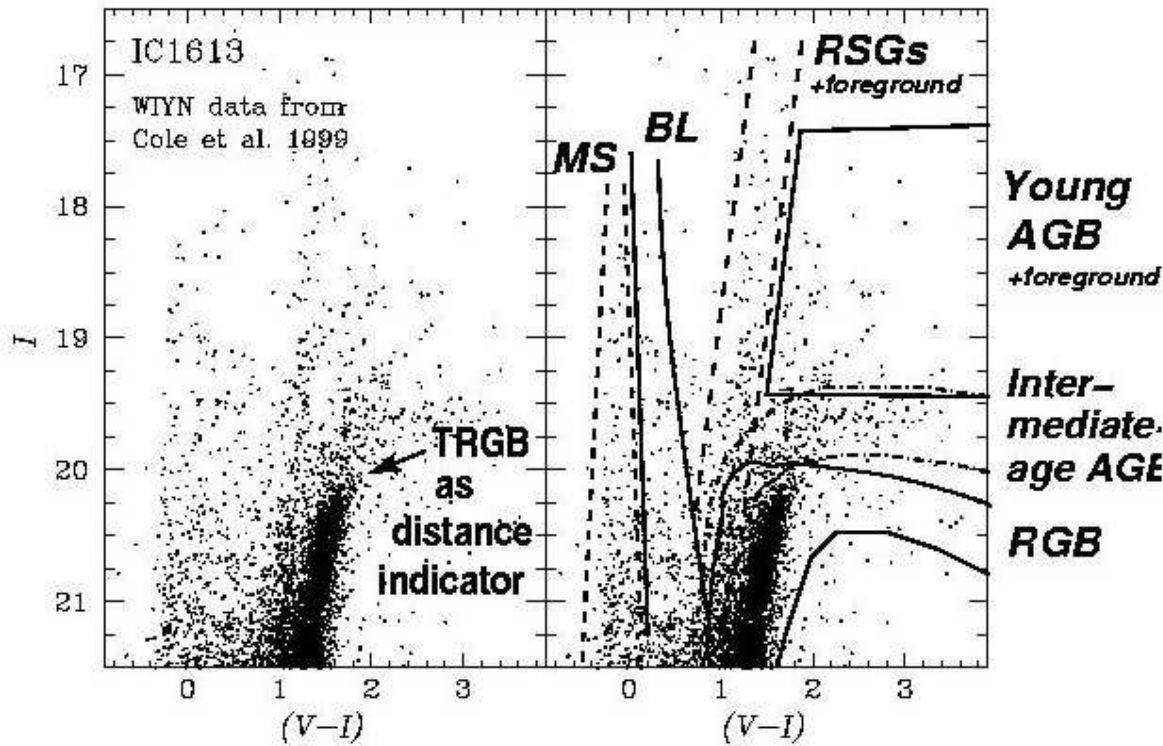, width=16.0cm}
\vskip-0.5cm
\caption{Color-magnitude diagram (CMD) 
of the upper portion of the red giant branch of the Local Group dIrr
IC 1613 (courtesy of Andrew Cole, see Cole et al.\ 1999) to illustrate the
information content of shallow 
CMDs of dwarf galaxies beyond the Local Group.  The
young main sequence (MS) and He-burning blue loop stars (BL) trace star 
formation events during the past $\sim 10$ to $\sim 300$ Myr, while 
He-burning red supergiants and giants mark events ranging from $\sim 10$ 
Myr to $\sim 1$ Gyr.  Young
asymptotic giant branch (AGB) stars tend to have ages older than 25
Myr, while the intermediate AGB represents star formation older than $\sim
1$ Gyr.  The tip of the red giant branch (TRGB) can be used as a distance
indicator for old populations more metal-poor than [Fe/H] $= -0.7$ dex.
The RGB slope and width yields estimates of the mean metallicity of the
old population and of its metallicity spread.  Depending on the mix of
populations, age-metallicity degeneracy effects may constrain the 
interpretation.}
\end{figure}

\section{Dwarf galaxies in the Local Volume}

Nearby dwarf galaxies outside of the Local Group have been studied in much
less detail.  We are currently carrying out an HST WFPC2 snapshot survey
to obtain upper red giant branch photometry of 200 nearby dwarf galaxies, and 
supporting ground-based observations to obtain surface photometry,
spectroscopic abundances, and velocities (Grebel et al.\ 2000).  At the 
same time, we are working
on extending the census of dwarf and low-surface-brightness galaxies 
to fainter magnitudes using the Sloan Digital Sky Survey.  

Our HST snapshot program yields information about the recent star formation
history and qualitative information about intermediate-age populations.  It
allows us to determine distances based on the tip of the red giant branch.
The slope of the red giant branch can be used for photometric metallicity
estimates (see Figure 1 for a schematic color-magnitude diagram).  This allows
us to derive both projected distances to the target galaxies and deprojected
distances between galaxies within galaxy groups.  We can thus study the
morphology--density relation in galaxy groups in three dimensions rather
than only in projection.  Studies using projected distributions of dwarf
galaxies in nearby groups suggest that dSphs and dEs tend to lie 
near massive galaxies, whereas dIrrs appear to be more widely distributed
(e.g., Jerjen, Binggeli, \& Freeman 2000).  
In Figure 2a  we show the the {\it deprojected} 
distances of dwarf galaxies around the Milky Way, M31, and M81.
The majority of these dwarfs are dSphs and dEs. 
In all three cases, we see the qualitatively similar tendency
for the dSphs and dEs to concentrate around the closest massive galaxy.

Considering our knowledge of the recent star formation history (based on
the observed presence or absence of young and intermediate-age populations) 
and of galaxy metallicity (through photometry or spectroscopy)
we can furthermore study the influence of environment on galaxy evolution.
Moreover, we can test the validity of global relationships such as the
metallicity--surface brightness--luminosity relation (first illustrated
for oxygen abundances and luminosities by Skillman, Kennicutt, \& Hodge
1989).
In Figure 2b this relation is shown for
dwarf galaxies in the Local Group and for dSphs in the M81 group.  The
mean metallicities for the M81 dSphs are at present purely based on photometry.
In spite of the significant scatter, dwarf galaxies in both groups show 
the expected positive
correlation between luminosity, surface brightness, and metallicity.

\begin{figure}[htb]
\epsfig{file=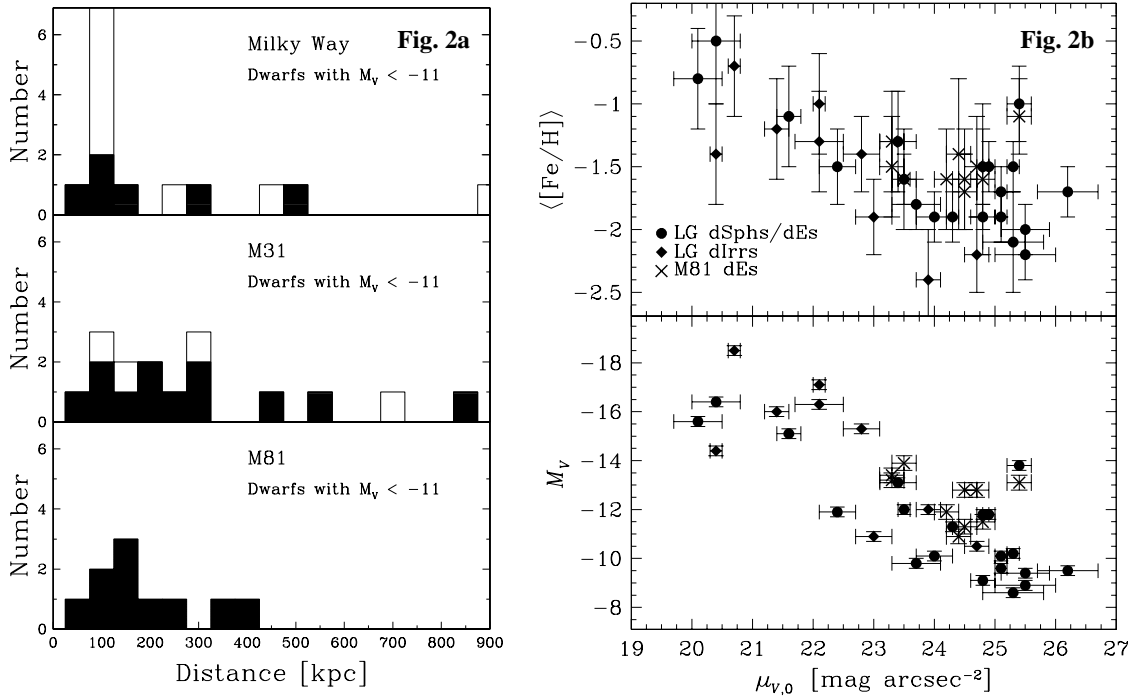, width=16.0cm}
\vskip-1.0cm
\caption{{\, \it Left panel: } The number of dwarf galaxies as a function
of distance from the Milky Way (upper left panel), from M31 (central left
panel), and from M81 (lower left panel).  Filled histogram bars stand for dwarf
galaxies with $M_V < -11$ mag in order to facilitate the comparison with the
M81 surroundings, since for M81 we only have data for galaxies
more luminous than this absolute $V$ magnitude.  The unfilled histograms
denote dwarf galaxies fainter than this absolute magnitude. The majority of 
the dwarf galaxies in the left panel are dSphs and dEs.  In the case of
M81, only dSphs are plotted.  These galaxies show a clear concentration
toward massive primaries.  {\, \it Right panel: }
Dwarf galaxies in the Local Group and in the M81 Group all tend to show
a positive correlation between mean metallicity, absolute magnitude, and
central surface brightness although the scatter 
exceeds the formal uncertainties of the individual measurements.  The data for 
the Local Group dwarfs were taken from Grebel (2000; note that the metallicity
error bars here are intended to represent metallicity spread rather than
uncertainty).  The data for dwarf
galaxies in the M81 Group are from Caldwell et al.\ (1998) and Karachentsev
et al.\ (2000b; 2001).  
}
\end{figure}

That dwarf galaxies show common trends for their integrated properties
in spite of their individual differences and irrespective of environment
may indicate that galaxy mass
is a determining factor in dwarf galaxy evolution.
On the other hand, the morphology--density relation illustrates that
environment is an important factor.  Clearly,
many important questions require further research,
including the mechanisms governing dwarf galaxy evolution, the origin of
the morphology--density relation, delayed evolution versus an ancient
common epoch of early star formation, gas loss mechanisms, metal loss
versus metal retention, enrichment as a function of time,
presence and nature of dark matter, and satellite orbits,
the uniqueness or generality of the Local Group, and differences between
groups of galaxies.  We are only at the beginning of the exploration of
the properties and evolution of the most frequent type of galaxy, the
dwarf galaxies.  The new era of large telescopes and a multi-wavelength
approach promises to help us
solve many of these fundamental questions.

{\small
\begin{description}{} \itemsep=0pt \parsep=0pt \parskip=0pt \labelsep=0pt
\item {\bf References}
\item Armandroff, T.~E., Davies, J.~E., \& Jacoby, G.~H. 1998, AJ, 116, 2287
\item Armandroff, T.~E., Jacoby, G.~H., \& Davies, J.~E. 1999, AJ, 118, 1220
\item Caldwell, N., Armandroff, T.~E., Da Costa, G.~S., \& Seitzer, P.
1998, AJ, 115, 535
\item Cole, A.~A., et al. 1999, AJ, 118, 1657
\item Cook, K.~H., Mateo, M., Olszewski, E.~W., Vogt, S.~S., Stubbs, C., \&
Diercks, A. 1999, PASP, 111, 306
\item Courteau, S., \& van den Bergh, S. 1999, AJ, 118, 337
\item Duc, P.-A., \& Mirabel, I.~F. 1998, A\&A, 333, 813
\item Gallart, C., Mart\'{\i}nez-Delgado, D., G\'omez-Flechoso, M.~A., \&
Mateo, M. 2001, AJ, 121, 2572
\item Grebel, E.~K. 1997, RvMA, 10, 29
\item Grebel, E.~K. 2000, Star Formation from the Small to the Large Scale,
33rd ESLAB Symposium, SP-445, eds.\ F.\ Favata, A.A.\ Kaas, \& A.\
Wilson (Noordwijk: ESA), 87
\item Grebel, E.~K., \& Stetson, P.~B. 1999, The Stellar Content of the Local 
Group, IAU Symp. 192, eds. P. Whitelock \& R. Cannon (ASP: Provo), 165
\item Grebel, E.~K., Seitzer, P., Dolphin, A.~E., Geisler, D., Guhathakurta,
P., Hodge, P.~W., Karachentsev, I.~D., Karachentseva, V.~E., \& Sarajedini, A.
2000, Stars, gas, and dust in galaxies: Exploring the links, ASP Conf. Ser.
Vol. 221, eds.\ D. Alloin, K. Olsen, \& C. Galaz (ASP: Provo), 147
\item Holtzman, J.~A., Smith, G.~H., \& Grillmair, C. 2000, AJ, 120, 3060
\item Hunter, D.~A. 1997, PASP, 109, 937
\item Hunter, D.~A., Hunsberger, S.~D., \& Roye, E.~W. 2000, ApJ, 542, 137
\item Jerjen, H., Binggeli, B., \& Freeman, K.~C. 2000, AJ, 119, 593
\item Karachentsev, I.~D., \& Karachentseva, V.~E. 1999, A\&A, 341, 335
\item Karachentsev, I.~D., Karachentseva, V.~E., Suchkov, A.~A., \& Grebel, 
E.~K. 2000a, A\&AS, 145, 415
\item Karachentsev, I.~D., Karachentseva, V.~E., Dolphin, A.~E., Geisler, D., 
Grebel, E.~K., Guhathakurta, P., Hodge, P.~W., Sarajedini, A., Seitzer, P.,
\& Sharina, M.~E. 2000b, A\&A, 363, 117 
\item Karachentsev, I.~D., Sharina, M.~E., Dolphin, A.~E., Geisler, D., 
Grebel, E.~K., Guhathakurta, P., Hodge, P.~W., Karachentseva, V.~E., 
Sarajedini, A., \& Seitzer, P. 2001, A\&A, in press
\item Lynden-Bell, D., \&  Lynden-Bell, R.~M. 1995, MNRAS, 275, 429
\item Lynds, R., Tolstoy, E., O'Neil, E.~J., \& Hunter, D.~A. 1998, AJ, 116, 146
\item Klypin, A., Kravtsov, A.~V., Valenzuela, O., \& Prada, F. 1999, ApJ, 522, 82
\item Mac Low, M.-M., \& Ferrara, A. 1999, ApJ, 513, 142
\item Majewski, S.~R. 1994, ApJ, 431, L17
\item Mateo, M. 1998, ARA\&A, 36, 435
\item Mayer, L., Governato, F., Colpi, M., Moore, B., Quinn, T., Wadsley, J.,
Stadel, J., \& Lake, G. 2001, ApJ, 547, L123
\item Matthews, L.~D., \& Gallagher, J.~S. 1997, AJ, 114, 1899
\item McGaugh, S.~S. 1994, ApJ, 426, 135
\item Minniti, D., \&  Zijlstra, A.~A. 1996, ApJ, 467, L13
\item Minniti, D., Zijlstra, A.~A., \& Alonso, M.~V. 1999, AJ, 117, 881
\item Sage, L.~J., Welch, G.~A., Mitchell, G.~F. 1998, ApJ, 507, 726
\item Schombert, J.~M., Pildis, R.~A., Eder, J.~A., \& Oemler, A. 1995, AJ, 110, 2067
\item S\'ersic, J.~L. 1968, Atlas de Galaxias Australes, Obs. Astron. de C\'ordoba
\item Skillman, E.~D., Kennicutt, R.~C., \& Hodge, P.~W. 1989, ApJ, 347, 875
\item van den Bergh, S. 1994,  ApJ, 428, 617
\item van Zee, L., Skillman, E.~D., \& Salzer, J.~J. 1998, AJ, 116, 1186
\item Welch, G.~A., Sage, L.~J., Mitchell, G.~F. 1998, ApJ, 499, 209
\item Wilcots, E.~M., \& Miller, B.~W. 1998, AJ, 116, 2363
\item Whiting, A.~B., Hau, G.~K.~T., \& Irwin, M. 1999, AJ, 118, 2767
\item Young, L.~M. 1999, AJ, 117, 1758
\item Young, L.~M., \& Lo, K.~Y. 1997a, ApJ, 490, 710
\item Young, L.~M., \& Lo, K.~Y. 1997b, ApJ, 476, 127
\end{description}
}

\end{document}